\documentclass[english]{article}
\usepackage[T1]{fontenc}
\usepackage[latin9]{inputenc}
\usepackage{color}
\usepackage{units}
\usepackage{amsmath}
\usepackage{setspace}
\usepackage{esint}

\makeatletter
\newcommand{\lyxaddress}[1]{
\par {\raggedright #1
\vspace{1.4em}
\noindent\par}
}

\makeatother

\usepackage{babel}
\begin{document}

\title{Heat flux in the presence of a gravitational field in a simple dilute
fluid: an approach based in general relativistic kinetic theory to
first order in the gradients.}

\author{D. Brun-Battistini$^{1}$, A. Sandoval-Villalbazo$^{1}$, A. L. Garcia-Perciante$^{2}$}

\maketitle

\lyxaddress{$^{1}$Depto. de Fisica y Matematicas, Universidad Iberoamericana
Ciudad de Mexico, Prolongacion Paseo de la Reforma 880, Mexico D.
F. 01219, Mexico.}

\lyxaddress{$^{2}$Depto. de Matematicas Aplicadas y Sistemas, Universidad Autonoma
Metropolitana-Cuajimalpa, Prol. Vasco de Quiroga 4871, Mexico D.F
05348, Mexico.}
\begin{abstract}
Richard C. Tolman analyzed the relation between a temperature gradient
and a gravitational field in an equilibrium situation. In 2012, Tolman\textquoteright s
law was generalized to a non-equilibrium situation for a simple dilute
relativistic fluid. The result in that scenario, obtained by introducing
the gravitational force through the molecular acceleration, couples
the heat flux with the metric coefficients and the gradients of the
state variables. In the present paper it is shown, by \textquotedblleft suppressing\textquotedblright{}
the molecular acceleration in Boltzmann\textquoteright s equation,
that a gravitational field drives a heat flux. This procedure corresponds
to the description of particle motion through geodesics, in which
a Newtonian limit to the Schwarzschild metric is assumed. The effect
vanishes in the non-relativistic regime, as evidenced by the direct
evaluation of the corresponding limit. 
\end{abstract}

\section{Introduction}

The problem of calculating the heat flux in a simple dilute relativistic
fluid due to a gravitational field, can be approached from different
perspectives. In 1930 Richard C. Tolman considered such system in
an equilibrium situation and showed that a gravitational field can
balance a temperature gradient, leading to a vanishing heat flux;
this is known as \textit{Tolman's law }\cite{Tolman original}\textit{.}
Several decades later, in 2012, an expression for the heat flux in
the presence of a linearized gravitational field, was established
in a non-equilibrium situation and Tolman's law was recovered when
heat flux vanishes and the equilibrium limit is attained \cite{Tolman PRD}\textit{.} 

On the other hand, in Ref. \cite{Kremer Sch} the heat flux is calculated
in a Schwarzschild isotropic metric in the context of general relativistic
kinetic theory concluding that the contribution of the gravitational
field vanishes. In that work it is argued that the effect obtained
in Ref. \cite{Tolman PRD} may arise from a metric factor not considered
in the equilibrium distribution function.

In the present paper it is shown that the gravitational contribution
to the heat flux does survive since the absolute molecular acceleration
term should not be included in Boltzmann's equation for structureless
particles. Moreover, the thermodynamic forces corresponding to the
gravitational field are associated to the covariant derivatives present
in the formalism. These important conceptual, as well as mathematical,
features improve the formalism presented in Ref. \cite{Tolman PRD}.

It is relevant to notice that another approach to this problem was
presented in 1984 by Wodarzik \cite{Wodarzik}. In that work the heat
flux is expressed using the corresponding Eckart constitutive equation
\cite{Eckart1-1}, and the well known hydrodynamic generic instabilities
that arise if this type of coupling is assumed are obtained \cite{HL}.
On the other hand, in the present paper we introduce the formal kinetic
theory definition of heat flux in the Navier-Stokes regime with no
acceleration coupling, instead particle number density, temperature
and gravitational potential gradients are present in the final expression,
\textcolor{black}{according to the assumptions of linear irreversible
thermodynamics.} Moreover, Wodarzik assumes that the \textit{fluid
bulk} moves following geodesics, argument that can be questioned as
individual molecules follow geodesics \cite{Weinberg}, but stresses
\textcolor{black}{deviate} the bulk from this type of dynamics. It
is well known that the hydrodynamic velocity $\left(\mathcal{U^{\nu}}\right)$
satisfies Euler equation \cite{ck,ChCow}:

\begin{equation}
\tilde{\rho}\dot{\mathcal{U^{\nu}}}=-h^{\nu\alpha}p_{,\alpha},\label{eq:1-1}
\end{equation}
where $\tilde{\rho}$ is $\left(\frac{n\varepsilon}{c^{2}}+\nicefrac{p}{c^{2}}\right)$,
with $n$ the particle number density, $\varepsilon$ the internal
energy, $p$ the local pressure and $c$ the speed of light; $h^{\nu\alpha}$
is the spatial projector defined as $g^{\nu\alpha}+\frac{\mathcal{U^{\nu}\mathcal{U^{\alpha}}}}{c^{2}}$,
where $p_{,\alpha}$ is the local pressure gradient and $g_{\mu\nu}$
is the metric tensor.

In order to establish the heat flux following the approach described
above, the rest of this paper is divided as follows: in Section II
we review a few concepts regarding the Boltzmann equation in the general
relativistic regime. In Section III, the term $\xi f^{(1)}$ is obtained,
following the general relativistic thermodynamic formalism in the
BGK approximation within Chapman-Enskog's expansion. The calculation
of the heat flux is shown in Section IV and Section V is devoted to
final remarks.

\section{Basic formalism: relativistic fluids}

\subsection{Basic elements of general relativity and the Schwarzschild metric}

\noindent A simple, non-degenerate gas is considered in a Riemaniann
space where the line element (arc length) is generally expressed as
\cite{Weinberg,Kremer}: 

\begin{equation}
ds^{2}=g_{\mu\nu}dx^{\mu}dx^{\nu},\label{eq:2-1}
\end{equation}
 where $g_{\mu\nu}$, as defined in Section I, is the metric tensor
which is essential when measuring distances in curved space-time.
Different metrics may be obtained from Einstein's field equations,
depending on the phenomenon to be analyzed. Here, a Schwarzschild
metric is used, with the property of being spherically symmetric and
static. A signature (1, 1, 1, -1) is considered so that $\mathcal{U^{\nu}\mathcal{U_{\nu}}}=-c^{2}$.
In such metric, the line element is given by:

\begin{equation}
ds^{2}=\frac{1}{\left(1-\frac{2GM}{c^{2}\tilde{r}}\right)}d\tilde{r}^{2}+\tilde{r}^{2}\left(d\theta^{2}+sin^{2}\theta d\varphi^{2}\right)-\left(1-\frac{2GM}{c^{2}\tilde{r}}\right)\left(dx^{4}\right)^{2},\label{eq:2-1-1}
\end{equation}
with $G$ being the gravitational constant, $c$ the speed of light,
$M$ the total mass, source of the gravitational field, and $r$,
$\theta$, $\varphi$ the spherical coordinates. As in Ref. \cite{Kremer Sch},
an isotropic Schwarzschild metric is used \cite{isotropic 2}, for
which the substitution $\tilde{r}=r\left(1+\frac{GM}{2c^{2}r}\right)$
is introduced in Eq.(\ref{eq:2-1-1}). Considering $\Phi\left(r\right)=\frac{GM}{r}$,
the gravitational potential, the Newtonian limit of the Schwarzschild
metric, which corresponds to weak field approximation $\left(\frac{\Phi}{c^{2}}\ll1\right)$
reads:
\begin{equation}
\tilde{\eta}_{\mu\nu}=\left(\begin{array}{cccc}
\left(1+\frac{2\Phi}{c^{2}}\right) & 0 & 0 & 0\\
0 & \left(1+\frac{2\Phi}{c^{2}}\right) & 0 & 0\\
0 & 0 & \left(1+\frac{2\Phi}{c^{2}}\right) & 0\\
0 & 0 & 0 & -\left(1-\frac{2\Phi}{c^{2}}\right)
\end{array}\right).\label{Schwarzschild metric}
\end{equation}

In order to perform the relevant calculations in the next sections
it is convenient to recall that in general relativity the velocity
is the derivative of the position four-vector with respect to the
arc length, so that the molecule's four-velocity is expressed as:

\begin{singlespace}
\begin{equation}
v^{\mu}=c\frac{dx^{\mu}}{ds}=\frac{cdx^{\mu}}{c\sqrt{g_{44}}d\tau}=\frac{1}{\sqrt{g_{44}}}\frac{dx^{\mu}}{d\tau}.\label{eq:2-3}
\end{equation}
Also, in the next section use of Christoffel symbols is made, with
the usual definition being:

\begin{equation}
\Gamma_{\alpha\beta}^{\mu}=\frac{g^{\mu\nu}}{2}\left(\frac{\partial g_{\alpha\nu}}{\partial x^{\beta}}+\frac{\partial g_{\beta\nu}}{\partial x^{\alpha}}-\frac{\partial g_{\alpha\beta}}{\partial x^{\nu}}\right)\label{eq:2-3-1}
\end{equation}
It must be noticed that the only non-vanishing terms for this metric
are the following:

\begin{equation}
\Gamma_{11}^{1}=\Gamma_{21}^{2}=\Gamma_{31}^{3}=\Gamma_{12}^{2}=\Gamma_{13}^{3}=\frac{\Phi'}{c^{2}\left(1+\frac{2\Phi}{c^{2}}\right)},\label{Christoffel 1}
\end{equation}

\begin{equation}
\Gamma_{22}^{1}=\Gamma_{33}^{1}=\Gamma_{44}^{1}=-\frac{\Phi'}{c^{2}\left(1+\frac{2\Phi}{c^{2}}\right)},\label{Christoffel 1-1}
\end{equation}

\begin{equation}
\Gamma_{14}^{4}=\Gamma_{41}^{4}=\frac{\Phi'}{c^{2}\left(-1+\frac{2\Phi}{c^{2}}\right)}.\label{Christoffel 1-1-1}
\end{equation}

\end{singlespace}

\subsection{Boltzmann's general relativistic equation}

Boltzmann's equation describes the evolution of the single-particle
distribution function and is expressed in special relativity as follows
\cite{ck,ChCow}:

\begin{equation}
\dot{f}=J(ff'),\label{Boltzmann 1}
\end{equation}
where $J(ff')$ is the collisional kernel and, in the special relativistic
regime, a dot denotes the total proper time ($\tau$) derivative.
Since $f$ is a function of the position ($x^{\mu}$) and velocity
($v^{\mu}$) four-vectors, $\dot{f}$ may be written as:

\begin{equation}
\dot{f}=\frac{\partial f}{\partial x^{\mu}}\dot{x}^{\mu}+\frac{\partial f}{\partial v^{\mu}}\dot{v}^{\mu}.\label{Boltzmann 2}
\end{equation}
On the other hand, in a general relativistic scenario, $\dot{f}$
corresponds to the arc length derivative ($\dot{f}=\frac{df}{ds}$),
with $s$ being the arc length \cite{Weinberg}. Also in this approach,
we have $\dot{x}^{\mu}=\frac{dx^{\mu}}{ds}$ and $\dot{v}^{\mu}=v^{\alpha}v_{;\alpha}^{\mu}=v^{\alpha}\left(\frac{\partial v^{\mu}}{\partial x^{\alpha}}+\Gamma_{\alpha\beta}^{\mu}v^{\beta}\right)$.
In this paper particles are assumed to lack structure (they don't
rotate and they don't have internal degrees of freedom), so that,
as a consequence of the field equations, the acceleration $\frac{dv^{\mu}}{ds}$
is zero, and the molecules move following geodesics, i. e., the following
equation is satisfied: 
\begin{equation}
\frac{d^{2}x^{\mu}}{ds^{2}}+\Gamma_{\alpha\beta}^{\mu}\frac{dx^{\alpha}}{ds}\frac{dx^{\beta}}{ds}=0.\label{eq:3}
\end{equation}
The Boltzmann's equation can thus be written as:

\begin{equation}
v^{\mu}f_{,\mu}=\frac{\partial f}{\partial x^{\mu}}v^{\mu}=J(ff').\label{Boltzmann 2-2-1-1}
\end{equation}
The collisional kernel can be modeled in a simple form using the BGK
approximation \cite{BGK}, so that Eq. (\ref{Boltzmann 2-2-1-1})
becomes:

\begin{equation}
v^{\mu}f_{,\mu}=-\frac{f-f^{(0)}}{\tau_{c}},\label{Boltzmann 2-2-1}
\end{equation}
where $\tau_{c}$ is a relaxation time and $f^{(0)}$ the local equilibrium
distribution function.

In order to calculate the heat flux, the first order in the gradients
correction to the distribution function $\left(f^{(1)}\right)$ must
be obtained. Following the kinetic theory approach, Chapman-Enskog's
method is now used, such that a solution given by $f=f^{(0)}+f^{(1)}$
is assumed, keeping only the first order in the gradients terms. By
substituting this solution in Eq.(\ref{Boltzmann 2-2-1}) and after
simple algebraic manipulation, one obtains:

\begin{equation}
f^{(1)}=-\tau_{c}\frac{\partial f^{(0)}}{\partial x^{\mu}}v^{\mu}.\label{eq:epsilonefe1}
\end{equation}

The next step consists in introducing the functional hypothesis \cite{degroor-1}
by means of which the factor $\frac{\partial f^{(0)}}{\partial x^{\mu}}$
is written in terms of the state variables $n$ (the local particle
number density), $T$ (the local temperature) and $\mathcal{U^{\mu}}$
(the hydrodynamic fluid velocity):

\begin{equation}
\frac{\partial f^{(0)}}{\partial x^{\mu}}=\frac{\partial f^{(0)}}{\partial n}\frac{\partial n}{\partial x^{\mu}}+\frac{\partial f^{(0)}}{\partial T}\frac{\partial T}{\partial x^{\mu}}+\frac{\partial f^{(0)}}{\partial\mathcal{U^{\alpha}}}\mathcal{U}_{;\mu}^{\alpha}.\label{regla de la cadena}
\end{equation}
The covariant derivative $\mathcal{U}_{;\mu}^{\alpha}$ introduced
in Eq. (\ref{regla de la cadena}) is imperative to preserve its invariance.
It is interesting to notice that in the definition of the covariant
derivative: 

\begin{equation}
\mathcal{U}_{;\mu}^{\alpha}=\frac{\partial\mathcal{U^{\alpha}}}{\partial x^{\mu}}+\Gamma_{\mu\nu}^{\alpha}\mathcal{U^{\nu}},\label{derivada covariante}
\end{equation}
the second term will become a thermodynamic force.

The Jüttner (Maxwell-Boltzmann relativistic) function must be considered
in order to establish the derivatives in Eq. (\ref{regla de la cadena}):

\begin{equation}
f^{(0)}=\frac{n}{4\pi m^{2}ck_{B}TK_{2}(\nicefrac{1}{z})}e^{\mathcal{U^{\alpha}}v_{\alpha}/k_{B}T},\label{Juttner}
\end{equation}
where $K_{\ell}(\nicefrac{1}{z})$ the modified Bessel function of
the second kind of order $\ell$ and $z=k_{B}T/mc^{2}$ is the relativistic
parameter, with $k_{B}$ the Boltzmann constant. In order to be precise,
the distribution function should include a metric factor $\left(\frac{1}{\sqrt{g_{44}}}\right)$
in the exponential argument, as remarked in Ref.\cite{Kremer Sch},
associated with the molecular velocity. However, as $v_{\alpha}$
and $x^{\alpha}$ are mutual independent variables, the derivatives
with respect to $x^{\alpha}$ will not include that metric factor.
Moreover, the invariant element in the phase space will also have
a metric factor $\sqrt{g}$, with $g$ the metric determinant, but
this \textit{will not affect the main result of this paper}, as it
will be a general common factor in the resulting integral expression.

Substituting in Eq. (\ref{regla de la cadena}) the covariant derivative's
definition (Eq. \ref{derivada covariante}) and the $f^{(0)}$ partial
derivatives that are found in \cite{ck}, one obtains:

\begin{equation}
\frac{\partial f^{(0)}}{\partial x^{\mu}}=\frac{f^{(0)}}{n}\frac{\partial n}{\partial x^{\mu}}+\left(1-\frac{\gamma_{(k)}}{z}-\frac{\mathcal{G}(\nicefrac{1}{z})}{z}\right)\frac{\partial T}{\partial x^{\mu}}+\frac{v_{\alpha}f^{(0)}}{zc^{2}}\left(\frac{\partial\mathcal{U^{\alpha}}}{\partial x^{\mu}}+\Gamma_{\mu\nu}^{\alpha}\mathcal{U^{\nu}}\right),\label{regla de la cadena-1}
\end{equation}
where $\mathcal{G}(\nicefrac{1}{z})=\frac{K_{3}(\nicefrac{1}{z})}{K_{2}(\nicefrac{1}{z})}$.

\section{The field contribution to $f^{(1)}$}

The rest of of the calculations in this paper, focus on the field
contribution to the heat flux, using the third term on the right hand
side term of Eq. (\ref{regla de la cadena-1}). Thus, we define

\begin{equation}
f_{\mathcal{U}}^{(1)}=-\tau_{c}\frac{v_{\alpha}f^{(0)}}{zc^{2}}\left(\frac{\partial\mathcal{U^{\alpha}}}{\partial x^{\mu}}+\Gamma_{\mu\nu}^{\alpha}\mathcal{U^{\nu}}\right)v^{\mu}.\label{eq:epsilonefe1-2}
\end{equation}
In order to establish the thermodynamic flux, space and time components
are separated as follows (latin indexes runs up to 3 and greek ones
up to 4):

\begin{equation}
f_{\mathcal{U}}^{(1)}=-\frac{\tau_{c}f^{(0)}}{zc^{2}}\left(v_{\alpha}v^{\ell}\frac{\partial\mathcal{U^{\alpha}}}{\partial x^{\ell}}+v_{\alpha}v^{4}\frac{\partial\mathcal{U^{\alpha}}}{\partial x^{(4)}}+v_{\alpha}v^{\mu}\Gamma_{\mu\nu}^{\alpha}\mathcal{U^{\nu}}\right).\label{eq:epsilonefe1-5}
\end{equation}
Notice that Eq. (\ref{eq:1-1}) can be expressed as:

\begin{equation}
\dot{\mathcal{U^{\nu}}}=\mathcal{U^{\mu}}\mathcal{U}_{;\mu}^{\nu}=-\frac{h^{\mu\nu}p_{,\mu}}{\tilde{\rho}}\label{eq:1-1-1}
\end{equation}
and using the covariant derivative definition (Eq. \ref{derivada covariante}):

\begin{equation}
\dot{\mathcal{U^{\nu}}}=\mathcal{U^{\mu}}\left(\frac{\partial\mathcal{U^{\nu}}}{\partial x^{\mu}}+\Gamma_{\mu\beta}^{\nu}\mathcal{U^{\beta}}\right)=-\frac{h^{\mu\nu}p_{,\mu}}{\tilde{\rho}}.\label{derivada covariante-1}
\end{equation}
The separation of spatial and temporal components yields:

\begin{equation}
\dot{\mathcal{U^{\nu}}}=\left(\mathcal{U^{\ell}}\frac{\partial\mathcal{U^{\nu}}}{\partial x^{\ell}}+\mathcal{U}^{(4)}\frac{\partial\mathcal{U^{\nu}}}{\partial x^{(4)}}+\Gamma_{\mu\beta}^{\nu}\mathcal{U^{\beta}}\mathcal{U^{\mu}}\right)=-\frac{h^{\mu\nu}p_{,\mu}}{\tilde{\rho}},\label{derivada covariante-1-1}
\end{equation}
so that the second term in parenthesis turns out to be:

\begin{equation}
\mathcal{U}^{(4)}\frac{\partial\mathcal{U^{\nu}}}{\partial x^{(4)}}=-\frac{h^{\nu\alpha}p_{,\alpha}}{\tilde{\rho}}-\Gamma_{\mu\beta}^{\nu}\mathcal{U^{\beta}\mathcal{U^{\mu}}}-\mathcal{U^{\ell}}\frac{\partial\mathcal{U^{\nu}}}{\partial x^{\ell}}.\label{derivada covariante-1-1-1}
\end{equation}

The substitution of Eq. (\ref{derivada covariante-1-1-1}) in Eq.
(\ref{eq:epsilonefe1-5}) will allow us to write the time derivatives
in terms of first order spatial gradients via the Euler equations.
This is necessary to guarantee the existence of the Chapman-Enskog
solution \cite{Courant =000026 Hilbert}. Following such procedure
one obtains:

\[
f_{\mathcal{U}}^{(1)}=-\frac{\tau_{c}f^{(0)}}{zc^{2}}\left(v_{\alpha}v^{\ell}\frac{\partial\mathcal{U^{\alpha}}}{\partial x^{\ell}}+\frac{v_{\alpha}v^{(4)}}{\mathcal{U}^{(4)}}\left(-\frac{h^{\nu\alpha}p_{,\alpha}}{\tilde{\rho}}-\Gamma_{\mu\beta}^{\nu}\mathcal{U^{\beta}\mathcal{U^{\mu}}}-\mathcal{U^{\ell}}\frac{\partial\mathcal{U^{\nu}}}{\partial x^{\ell}}\right)\right)
\]

\begin{equation}
-\frac{\tau_{c}f^{(0)}}{zc^{2}}v_{\alpha}v^{\mu}\Gamma_{\mu\nu}^{\alpha}\mathcal{U^{\nu}}{\color{black})}.\label{eq:epsilonefe1-6-1}
\end{equation}
In what follows, only the terms depending on Christoffel symbols will
be taken into account since they contain the curvature, and thus gravitational
effects:

\begin{equation}
f_{[g]}^{(1)}=-\frac{\tau_{c}f^{(0)}}{zc^{2}}\left(\frac{-v_{\alpha}v^{(4)}}{\mathcal{U}^{(4)}}\Gamma_{\beta\lambda}^{\alpha}\mathcal{U^{\lambda}}\mathcal{U^{\beta}}+v_{\alpha}v^{\mu}\Gamma_{\mu\nu}^{\alpha}\mathcal{U^{\nu}}\right),\label{eq:epsilonefe1-7-1}
\end{equation}
and calculations will be performed in the comoving frame where $\mathcal{U^{\nu}}=(0,0,0,c)$,
$k^{\nu}\equiv(k^{1},k^{2},k^{3},c)$ and $v_{\eta}=\gamma_{(k)}k_{\eta}$
\textcolor{black}{with} $k_{\eta}$ representing the chaotic velocity
(the molecule's velocity measured in the comoving frame) \cite{ck,Alma y Ana}
and $\gamma_{(k)}=\left(1-\frac{k^{2}}{c^{2}}\right)^{-1/2}$. \textcolor{black}{With
these substitutions we obtain the following expression for the gravitational
contribution to the first order in the gradients distribution function:}

\begin{equation}
f_{[g]}^{(1)}=-\frac{\tau_{c}\gamma_{(k)}^{2}f^{(0)}}{zc}\left(k_{\alpha}k^{\mu}\Gamma_{\mu4}^{\alpha}-ck_{\alpha}\Gamma_{44}^{\alpha}\right).\label{eq:epsilonefe1-8}
\end{equation}
In Ref. \cite{Kremer Sch}, the terms corresponding to the ones in
brackets in Eq. (\ref{eq:epsilonefe1-8}) cancel out as the molecule
acceleration term is expressed in terms of a force induced by the
field, i. e., \textcolor{black}{$\frac{dv^{\mu}}{d\tau}=-\Gamma_{\alpha\beta}^{\mu}v^{\alpha}v^{\beta}$},
together with the missing covariant derivative of the hydrodynamic
velocity in Eq. (18) of Ref. \cite{Kremer Sch}. 

Before proceeding to the heat flux calculation it is useful to notice
that the summation $k_{\alpha}k^{\mu}\Gamma_{\mu4}^{\alpha}$vanishes,
so that:

\begin{equation}
f_{[g]}^{(1)}=\frac{\tau_{c}\gamma_{(k)}^{2}f^{(0)}}{z}\left(k_{\alpha}\Gamma_{44}^{\alpha}\right).\label{eq:epsilonefe1-8-1}
\end{equation}
Equation (\ref{eq:epsilonefe1-8-1}) is the basis of the heat flux
field contribution that will be calculated in the next section.

\section{Heat flux calculation with a spherically symmetric static metric}

Kinetic theory's definition of heat flux in the comoving frame is
expressed as follows \cite{Jnet 12}:
\begin{equation}
J_{[Q]}^{\ell}=mc^{2}\int k^{\ell}f^{(1)}\gamma_{(k)}\left(\gamma_{(k)}-1\right)d^{*}\!K,\label{eq:heat flux}
\end{equation}
where the volume element is $d^{*}\!K=4\pi c^{3}(\gamma_{(k)}^{2}-1)^{\nicefrac{1}{2}}d\gamma$
\cite{Liboff,Alma y Ana}. Now, using the expression (\ref{eq:epsilonefe1-8-1})
for $f_{[g]}^{(1)}$, the gravitational field contribution to the
hat flux can be written as:

\begin{equation}
J_{[Qg]}^{\ell}=\frac{\tau_{c}mc}{z}\int k^{\ell}f^{(0)}\gamma_{(k)}^{3}\left(\gamma_{(k)}-1\right)\left(ck_{\alpha}\Gamma_{44}^{\alpha}\right)d^{*}\!K\label{eq:heat flux-1}
\end{equation}
Performing the calculations (see the appendix for details) it is obtained
that:

\begin{equation}
J_{[Qg]}^{\ell}=\frac{\tau_{c}mnc^{4}}{3z}\tilde{\eta}^{\ell\ell}\Gamma_{44}^{\ell}\left[1+5z\mathcal{G}(\nicefrac{1}{z})-\mathcal{G}(\nicefrac{1}{z})\right]\label{eq:heat flux-3-5}
\end{equation}
In a Schwarzschild metric, after substitution of the Christoffel symbols,
\textcolor{black}{it is found that the gravitational contribution
to the heat flux in a simple dilute general relativistic fluid is:}

\begin{equation}
J_{[Qg]}^{\ell}=nm\tau_{c}c^{4}\left[1+5z\mathcal{G}(\nicefrac{1}{z})-\mathcal{G}(\nicefrac{1}{z})\right]\frac{\Phi^{,\ell}}{c^{2}},\label{eq:heat flux-3-6}
\end{equation}
This effect, that is not present in the non-relativistic case, does
not vanish in the relativistic regime, neither in special nor in general
relativity, at least in the case of a static and symmetrical metric.
When the limit when \textcolor{black}{$z\rightarrow0$ (non-relativistic
regime) is considered one obtains:}

\begin{equation}
J_{[Qg]}^{\ell}=\frac{n\tau_{c}k_{B}T}{z}\Phi^{,\ell}.\label{eq:heat flux-3-6-2}
\end{equation}
This result is in total agreement with the one presented in Refs.
\cite{Tolman original,Tolman PRD}.

\section{Final remarks}

It was shown that a gravitational field has a contribution to the
heat flux of a simple dilute general relativistic fluid. A static
and symmetric metric was assumed during the calculation. Such result
is a consequence of using the covariant derivative in Boltzmann's
equation for structureless particles that follow geodesic trajectories.
The expression obtained leads to the non-relativistic result of the
heat flux being coupled solely to the temperature gradient in the
corresponding limit, as expected. In the case of flat spacetime and
in cartesian coordinates, the covariant derivative vanishes (Christoffel
symbols are zero) and the field coupling with the heat flux survives
only if a linearized gravity approach is taken into account \cite{Tolman PRD}. 

In contrast with related works \cite{Kremer Sch}, the gravitational
field and the particle number density are both coupled with heat flux
and will contribute to the entropy production; in other words, the
curvature of spacetime itself contributes to the heat flux and produces
entropy. A complete evaluation of entropy production \textcolor{black}{($\sigma=-\frac{k_{B}}{c}\int J(ff')\varphi d^{*}\!v$},
with \textcolor{black}{$\varphi=\frac{\varsigma f^{(1)}}{f^{(0)}}$)}
will be addressed in a separate work. 

The next step corresponds to the study of tensor effects, which involve
viscosity coefficients and the use of Christoffel symbols as thermodynamic
forces. In the longer term a similar formalism will be presented to
thoroughly analyze the entropy production associated with the field
itself.\bigskip{}

\textsf{\textbf{\textit{\Large{}Acknowledgments}}}{\Large \par}

The authors wish to thank MSc. Alma Sagaceta-Mejia and Dr. Humberto
Mondragon-Suarez for their valuable comments to this work and acknowledge
support from CONACyT through grant number CB2011/167563. 

\bigskip{}

\textsf{\textbf{\textit{\Large{}Appendix }}}{\Large \par}

In this appendix, details of the gravitational field contribution
to the heat flux presented in Section IV are described. Performing
the summation over \textcolor{black}{$\alpha$} in Eq. (\ref{eq:heat flux-1})
we have:

\begin{equation}
J_{[Qg]}^{\ell}=\frac{\tau_{c}mc^{2}}{z}\int k^{\ell}f^{(0)}\gamma_{(k)}^{3}\left(\gamma_{(k)}-1\right)\left(k_{1}\Gamma_{44}^{1}+k_{2}\Gamma_{44}^{2}+k_{3}\Gamma_{44}^{3}+k_{4}\Gamma_{44}^{4}\right)d^{*}\!K,\label{eq:heat flux-2}
\end{equation}
which leads to \textcolor{black}{: }

\begin{equation}
J_{[Qg]}^{\ell}=\frac{\tau_{c}mc^{2}}{z}\int f^{(0)}\gamma_{(k)}^{3}\left(\gamma_{(k)}-1\right)\varGamma d^{*}\!K.\label{eq:heat flux-3-1-1}
\end{equation}
where factor \textbf{\textcolor{black}{${\color{black}{\color{red}{\color{black}{\color{black}\varGamma}}}}$
}}is:

\begin{equation}
{\color{black}{\color{red}{\color{black}{\color{black}\varGamma}}}}=\left(\begin{array}{c}
\Gamma_{44}^{1}k^{(1)}k_{(1)}+\Gamma_{44}^{2}k^{(1)}k_{(2)}+\Gamma_{44}^{3}k^{(1)}k_{(3)}+\Gamma_{44}^{4}k^{(1)}k_{(4)}\\
\Gamma_{44}^{1}k^{(2)}k_{(1)}+\Gamma_{44}^{2}k^{(2)}k_{(2)}+\Gamma_{44}^{3}k^{(2)}k_{(3)}+\Gamma_{44}^{4}k^{(2)}k_{(4)}\\
\Gamma_{44}^{1}k^{(3)}k_{(1)}+\Gamma_{44}^{2}k^{(3)}k_{(2)}+\Gamma_{44}^{3}k^{(3)}k_{(3)}+\Gamma_{44}^{4}k^{(3)}k_{(4)}
\end{array}\right).\label{eq:heat flux-3-1-1-2}
\end{equation}

\textcolor{black}{In order to write Eq.(\ref{eq:heat flux-3-1-1})
in covariant form, it will be used that $\tilde{\eta}^{\ell\ell}k_{(\ell)}=k^{(\ell)}$
for $\ell=1,2,3$, and thus $k^{(1)}k_{(1)}=\left(k^{(1)}\right)^{2}$.
Taking into account that }all the terms with the factor $k^{(\mu)}k_{(\nu)}$for
$\mu\neq\nu$ vanish for parity\textcolor{black}{{} reasons, we shall
have:}

\begin{equation}
J_{[Qg]}^{\ell}=\frac{\tau_{c}mc^{2}}{z}\int f^{(0)}\gamma_{(k)}^{3}\left(\gamma_{(k)}-1\right)\left(\begin{array}{c}
\Gamma_{44}^{1}\tilde{\eta}^{11}\left[k^{(1)}\right]^{2}\\
\Gamma_{44}^{2}\tilde{\eta}^{22}\left[k^{(2)}\right]^{2}\\
\Gamma_{44}^{3}\tilde{\eta}^{33}\left[k^{(3)}\right]^{2}
\end{array}\right)d^{*}\!K.\label{eq:heat flux-3-1-1-1}
\end{equation}
Since the three integrals are equal and $\left[k^{(1)}\right]^{2}+\left[k^{(2)}\right]^{2}+\left[k^{(3)}\right]^{2}=k^{2}$,
Eq. (\ref{eq:heat flux-3-1-1-1}) can be written as:

\begin{equation}
J_{[Qg]}^{\ell}=\frac{\tau_{c}mc^{2}}{3z}\intop f^{(0)}\gamma_{(k)}^{3}\left(\gamma_{(k)}-1\right)k^{2}\tilde{\eta}^{\ell\ell}\Gamma_{44}^{\ell}d^{*}\!K.\label{eq:heat flux-3-1-1-1-1}
\end{equation}
The next step is to express the integral in terms of \textcolor{black}{$\gamma_{(k)}$:}

\begin{equation}
J_{[Qg]}^{\ell}=\frac{\tau_{c}mc^{2}}{3z}\int_{1}^{\infty}f^{(0)}\gamma_{(k)}^{3}\left(\gamma_{(k)}-1\right)\frac{c^{2}(\gamma_{(k)}^{2}-1)}{\gamma_{(k)}^{2}}\tilde{\eta}^{\ell\ell}\Gamma_{44}^{\ell}4\pi c^{3}(\gamma_{(k)}^{2}-1)^{\nicefrac{1}{2}}d\gamma_{(k)},\label{eq:heat flux-3-1-1-1-1-1}
\end{equation}
and after Juttner's distribution function is substituted we obtain
that:

\begin{equation}
J_{[Qg]}^{\ell}=\frac{\tau_{c}mnc^{4}}{3zK_{2}(\nicefrac{1}{z})}\tilde{\eta}^{\ell\ell}\Gamma_{44}^{\ell}\int_{1}^{\infty}e^{-\nicefrac{\gamma_{(k)}}{z}}\gamma_{(k)}\left(\gamma_{(k)}-1\right)(\gamma_{(k)}^{2}-1)^{\nicefrac{3}{2}}d\gamma_{(k)},\label{eq:heat flux-3-1-1-1-1-1-1}
\end{equation}
so that:

\begin{equation}
J_{[Qg]}^{\ell}=\tau_{c}nk_{B}T\left\{ 1+5z\mathcal{G}(\nicefrac{1}{z})-\mathcal{G}(\nicefrac{1}{z})\right\} \Phi^{,\ell},\label{4-1}
\end{equation}
from where the field contribution to the heat flux is given by Eq.(\ref{eq:heat flux-3-5}):

\[
J_{[Qg]}^{\ell}=\frac{\tau_{c}nmc^{2}}{z}\left[1+5z\mathcal{G}(\nicefrac{1}{z})-\mathcal{G}(\nicefrac{1}{z})\right]\Phi^{,\ell}.
\]
Here use has been made of $z=\frac{k_{B}T}{mc^{2}}$, the relativistic
parameter and the substitution of $\tilde{\eta}^{\ell\ell}\Gamma_{44}^{\ell}$
values has been performed.


\begin{thebibliography}{10}
\bibitem[1]{Tolman original}Tolman, R. C.; Physics. Rev. 35, 904-924
(1930); see also Phys. Rev. \textbf{36}, 1791-1798 (1930).

\bibitem[2]{Tolman PRD}Sandoval-Villalbazo, A., Garcia-Perciante,
A. L. and Brun-Battistini, D.; Phys. Rev. D \textbf{86}, 084015 (2012).

\bibitem[3]{Kremer Sch}Kremer, G. M.; Relativistic gas in a Schwarzschild
metric; J. Stat. Mech. P04016 (2013).

\bibitem[4]{Wodarzik}Wodarzik, U. F.; Phys. Rev. A \textbf{30}, 3
(1984).

\bibitem[5]{Eckart1-1} Eckart, C.; Phys. Rev. \textbf{58,} 267 (1940);
Phys. Rev. \textbf{58,} 919 (1940).

\bibitem[6]{HL}Hiscock, A. and Lindblom, L.; Phys. Rev. D \textbf{31,}
725 (1985).

\bibitem[7]{Weinberg}Weinberg, S.; \textit{Gravitation and Cosmology:
Principles and Applications of the General Theory of Relativity};
John Wiley \& Sons, N. Y. (1972).

\bibitem[8]{ck}Cercignani, C. and Kremer, G. M.; \textit{The relativistic
Boltzmann equation: theory and applications} (Birkhauser Verlag, Basel,
2002).

\bibitem[9]{ChCow}Chapman, S. and Cowling, T. G. \textit{; The mathematical
theory of non-uniform gases} (Cambridge Mathematical Library, United
Kingdom, 1971), 3rd. ed.

\bibitem[10]{Kremer} \textcolor{black}{Kremer, G. M.; The Boltzmann
equation in special and general relativity; 28th International Symposium
on Rarefied Gas Dynamics (Zaragoza, Spain, July 9-13, 2012). }

\bibitem[11]{isotropic 2}See for example Crothers, S. J.; Progress
in Physics, (2006) and Buchdahl, H.; International Journal of Theoretical
Physics, \textbf{24}, 7 (1985).

\bibitem[12]{BGK}Bhatnagar, P. L., Gross, E. P. \& Krook, M.; Phys.
Rev. \textbf{94}, 3 (1954). 

\bibitem[13]{degroor-1}de Groot, S. R., van Leeuwen, W. A. and van
der Wert, \textit{Ch.; Relativistic Kinetic Theory} (North Holland
Publ. Co., Amsterdam, 1980).

\bibitem[14]{Courant =000026 Hilbert}Courant, R. \& Hilbert, D.;
Methods of mathematical physics; John Wiley and Sons, vol. 1, New
York (1989). 

\bibitem[15]{Jnet 12}Garcia-Perciante, A. L., Sandoval-Villalbazo,
A. y Garcia-Colin, L. S.; Journal of Non-Equilibrium Thermodynamics,
Volume 37, Issue 1, Pages 43-61 (2012).

\bibitem[16]{Liboff}Liboff, R. L.; Kinetic Theory: Classical, Quantum
and Relaivistic Descriptions (Springer, New York, 2003), 3rd ed.

\bibitem[17]{Alma y Ana}Garcia-Perciante, A. L. and Mendez, A. R.;
Gen. Relativ. Gravit. \textbf{43}, 2257 (2011).\end{thebibliography}
\end{document}